\documentclass[prl,twocolumn,aps,showpacs,preprintnumbers,amsmath,amssymb, superscriptaddress]{revtex4-1}
\pdfoutput=1
\usepackage{bm}
\usepackage{graphicx}
\usepackage{color}
\begin{document}

\title{
%Electronic structure of multiband superconductor TlNi$_2$Se$_2$: a weakly correlated system with van Hove singularities near $E_F$\textcolor{red}{)}
Camelback-shaped band reconciles heavy electron behavior with weak electronic Coulomb correlations in superconducting TlNi$_2$Se$_2$}

\author{N. Xu}

\email{nan.xu@psi.ch}

\affiliation{Swiss Light Source, Paul Scherrer Insitut, CH-5232 Villigen PSI,
Switzerland}
\affiliation{Institute of Condensed Matter Physics, \'Ecole Polytechnique F\'ed\'erale de Lausanne, CH-1015 Lausanne, Switzerland}

\author{C. E. Matt}

\affiliation{Swiss Light Source, Paul Scherrer Insitut, CH-5232 Villigen PSI,
Switzerland}

\affiliation{Laboratory for Solid State Physics, ETH Z\"urich, CH-8093 Z\"urich,
Switzerland}

\author{P. Richard}
\affiliation{Beijing National Laboratory for Condensed Matter Physics, and Institute of Physics, Chinese Academy of Sciences, Beijing 100190, China}
\affiliation{Collaborative Innovation Center of Quantum Matter, Beijing, China}

\author{A. van Roekeghem}
\affiliation{Beijing National Laboratory for Condensed Matter Physics, and Institute of Physics, Chinese Academy of Sciences, Beijing 100190, China}
\affiliation{Centre de Physique Th{\'e}orique, Ecole Polytechnique, CNRS-UMR7644, 91128 Palaiseau, France}

\author{S. Biermann}
\affiliation{Centre de Physique Th{\'e}orique, Ecole Polytechnique, CNRS-UMR7644, 91128 Palaiseau, France}
\affiliation{Coll{\`e}ge de France - 11 place Marcelin Berthelot, 75005 Paris, France}
\affiliation{Kavli Institute for Theoretical Physics, University of California - Santa Barbara, CA 93106, USA}

\author{X. Shi}
\affiliation{Swiss Light Source, Paul Scherrer Insitut, CH-5232 Villigen PSI, Switzerland}
\affiliation{Beijing National Laboratory for Condensed Matter Physics, and Institute of Physics, Chinese Academy of Sciences, Beijing 100190, China}

\author{S.-F. Wu}
\affiliation{Beijing National Laboratory for Condensed Matter Physics, and Institute of Physics, Chinese Academy of Sciences, Beijing 100190, China}

\author{H. W. Liu}
\affiliation{Beijing National Laboratory for Condensed Matter Physics, and Institute of Physics, Chinese Academy of Sciences, Beijing 100190, China}

\author{D. Chen}

\author{T. Qian}
\affiliation{Beijing National Laboratory for Condensed Matter Physics, and Institute of Physics, Chinese Academy of Sciences, Beijing 100190, China}

\author{N. C. Plumb}
\affiliation{Swiss Light Source, Paul Scherrer Insitut, CH-5232 Villigen PSI, Switzerland}

\author{M. Radovi\'c}
\affiliation{Swiss Light Source, Paul Scherrer Insitut, CH-5232 Villigen PSI, Switzerland}
\affiliation{SwissFEL, Paul Scherrer Institut, CH-5232 Villigen PSI, Switzerland}

\author{Hangdong Wang}
\author{Qianhui Mao}
\author{Jianhua Du}
\affiliation{Department of Physics, Zhejiang University, Hangzhou 310027, China}
\author{Minghu Fang}

\affiliation{Department of Physics, Zhejiang University, Hangzhou 310027, China}
\affiliation{Collaborative Innovation Center of Advanced Microstructures, Nanjing University, Nanjing 210093, China}

\author{J. Mesot}
\affiliation{Swiss Light Source, Paul Scherrer Insitut, CH-5232 Villigen PSI, Switzerland}
\affiliation{Institute of Condensed Matter Physics, \'Ecole Polytechnique F\'ed\'erale de Lausanne, CH-1015 Lausanne, Switzerland}
\affiliation{Laboratory for Solid State Physics, ETH Z\"urich, CH-8093 Z\"urich, Switzerland}

\author{H. Ding}
\affiliation{Beijing National Laboratory for Condensed Matter Physics, and Institute of Physics, Chinese Academy of Sciences, Beijing 100190, China}
\affiliation{Collaborative Innovation Center of Quantum Matter, Beijing, China}

\author{M. Shi}
\email{ming.shi@psi.ch}
\affiliation{Swiss Light Source, Paul Scherrer Insitut, CH-5232 Villigen PSI, Switzerland}

\date{\today}

%\begin{minipage}[t]{6.8in}
\begin{abstract}
{Using high-resolution photoemission spectroscopy and first-principles calculations, we characterize superconducting TlNi$_2$Se$_2$ as a material with weak electronic Coulomb correlations leading to a bandwidth renormalization of 1.4. We identify a camelback-shaped band, whose energetic position strongly depends on the selenium height. While this feature is universal in transition metal pnictides, in TlNi$_2$Se$_2$ it lies in the immediate vicinity of the Fermi level, giving rise to a pronounced van Hove singularity. The resulting heavy band mass resolves the apparent puzzle of a large normal-state specific heat coefficient (Phys. Rev. Lett. 112, 207001) in this weakly correlated compound.}

\end{abstract}


%\end{minipage}
\maketitle
%\narrowtext

The discovery of high-temperature superconductivity in iron-chalcogenide A$_x$Fe$_{2-y}$Se$_2$ (A = Tl, K, Cs, Rb) \cite{JG_Guo_PRB2010, MHFang_EPL_122}  has raised a lot of attention because their unique Fermi surface (FS) topology without hole pocket \cite{Qian_PRL2011,Y_Zhang_NatureMat2011, D_MouPRL2011, XP_WangEPL2011, ZH_LiuPRL2012, XP_WangEPL2012} challenges the electron-hole quasi-nesting scenario as the main Cooper pairing force in the Fe-based superconducting materials \cite{Richard_RoPP2011}. However, due to the phase separation associated with Fe vacancy ordering  \cite{W_LiNatPhys8, Z_WangPRB83} in A$_x$Fe$_{2-y}$Se$_2$, it is still unclear whether superconductivity in this compound is related to that of the other iron-based superconductors. Very recently, it has been found that the iso-structural material TlNi$_2$Se$_2$ with two more electrons on the 3d shell shows superconductivity with $T_c$ = 3.7 K \cite{MHFang_TlNi2Se2_2013}. In contrast to Tl(K)$_x$Fe$_{2-y}$Se$_2$, X-ray diffraction results indicate that this system is homogeneous without Ni vacancy or phase separation, and stoichiometric TlNi$_{2}$Se$_2$ was confirmed by energy dispersive X-ray spectroscopy \cite{MHFang_TlNi2Se2_2013}. A large normal-state Sommerfeld coefficient has been attributed to heavy fermion behavior \cite{MHFang_TlNi2Se2_2013}, and its square root relationship with magnetic field in the mixed state suggests $d$-wave wave paring symmetry. On the other hand, thermal conductivity measurements suggest multiple nodeless superconducting gaps \cite{SYLi_TlNi2Se2_Thermo_conductivity} in TlNi$_2$Se$_2$. Therefore, it is crucial to determine the electronic structure of TlNi$_{2}$Se$_2$ to understand the nature of the reported heavy fermions in this material and to establish possible connections with other unconventional superconductors.

In this letter, we present high-resolution angle-resolved photoemission spectroscopy (ARPES) results on TlNi$_{2}$Se$_2$.
Wide range photon energy dependent measurements reveal that the electronic structure exhibits three dimensionality, with four bands crossing the Fermi level ($E_F$). Our density functional theory (DFT) calculations, renormalized by a factor of 1.4, match the experimentally determined band structure very well, indicating weaker correlation effects in TlNi$_{2}$Se$_2$ than in its cousin Tl(K)$_x$Fe$_{2-y}$Se$_2$ \cite{Qian_PRL2011}. 
%The FS of TlNi$_{2}$Se$_2$ shows very distinct topology which differs drastically from that of its Fe-chalcogenide counterpart. However, in first approximation, the band dispersion of TlNi$_2$Se$_2$ resembles that of Tl(K)Fe$_2$Se$_2$ with a chemical potential shifted by $\sim$ 0.5 eV, suggesting that TlNi$_2$Se$_2$ can be considered as a heavily electron doped Tl(K)Fe$_2$Se$_2$.
We reveal that the flat top of one of the bands %with $d_{xy}$ orbital character 
($\gamma$) exhibits an asymmetric back-bending feature resulting in a camelback shape near $E_F$ at $k$ = (0,0,$\pi$), the Z point of the first Brillouin zone (BZ). This shape is captured by our DFT calculations and seems to be a general feature of transition metal pnictides \cite{Andersen_Boeri_Calculation}. This $\gamma$ band forms four small FS lobes around the Z point, and between the lobes four flat parts resulting in a van Hove singularity (VHS) near $E_F$ are identified. This finding provides a natural explanation to the heavy electrons feature inferred from electronic specific heat and the upper critical field measurements \cite{MHFang_TlNi2Se2_2013} in this weakly correlated system \cite{compair}. 

%Indeed, in TlNi$_2$Se$_2$ $\gamma$ takes on a value of 40 mJ/mol K$^2$, which is 6 times larger than that of BaFe$_2$As$_2$ (5.6 mJ/mol K$^2$) with stronger correlations (renormalisation factor $\sim$ 3 \cite{RichardPRL2010}).

\begin{figure}[!t]
\begin{center}
\includegraphics[width=3.4in]{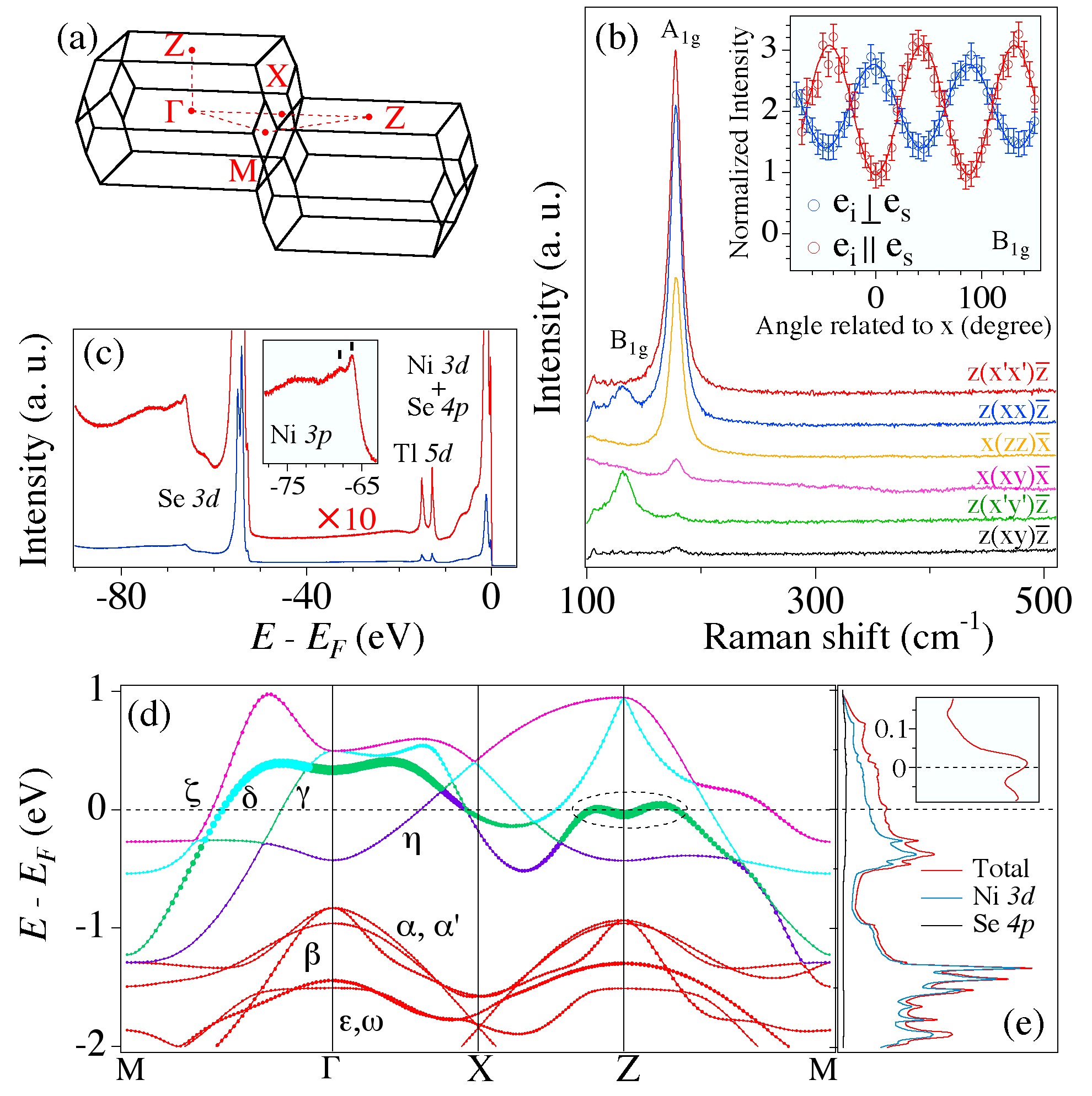}
%\epsfile{file={Fig_1},scale=0.8}
\end{center}
\caption{\label{Fig1_core}(Color online): (a)Two adjacent BZs of TlNi$_2$Se$_2$. (b) Raman spectra recorded under several polarization configurations \cite{raman_exp}. (c) Core level spectra of TlNi$_2$Se$_2$ recorded with 150 eV photons. The inset is a close up of the Ni $3p$ levels. (d) LDA band structure plotted along high symmetry lines. The line-width indicates the Se 4$p$ spectral weight. (e) Density of states from LDA calculations.}
\end{figure}

Large single crystals of TlNi$_{2}$Se$_2$ were grown by the self-flux method \cite{MHFang_TlNi2Se2_2013}. ARPES measurements were performed at SIS beamline of Swiss Light Source, and at the 1-cubed ARPES end-station of BESSY using VG-Scienta R4000 electron analyzers with photon energy ranging from 20 to 70 eV. The angular resolution was set to 0.2$^{\circ}$ and the energy resolution to 5$\sim$10 meV. Clean surfaces for the ARPES measurements were obtained by cleaving crystals \emph{in situ} in a working vacuum better than 5 $\times$ 10$^{-11}$ Torr. We label the momentum ($k$) values with respect to the 1 Fe/unit cell BZ, with the high symmetry points defined in Fig. \ref{Fig1_core}(a). The Fermi level of the samples was referenced to that of a gold film evaporated onto the sample holder. Raman data have been recorded at room temperature using a 514.5 nm laser source and the single mode of a Horiba Jobin Yvon-T64000 micro-Raman spectrometer equipped with a CCD camera. 

The Raman spectra in Fig. \ref{Fig1_core}(b) show one $A_{1g}$ mode (177.9 cm$^{-1}$) and one $B_{1g}$ (132.8 cm$^{-1}$) phonon, similar to the observation on the 122-ferropnictides \cite{AM_ZhangCPB22}, which share the same crystal structure. As illustrated in the inset, the intensity of these peaks are in perfect concordance with the four-fold symmetry of the crystal and the spectra are exempt of extra modes, which suggests that our TlNi$_2$Se$_2$ samples are not phase-separated and do not show Ni vacancy ordering, in agreement with a previous study \cite{Lazarevic_PRB87} on isostructural KNi$_2$Se$_2$.

%\begin{figure}[htbp]
%\begin{center}
%\epsfile{file=,scale=0.8}
%\caption{{\bf default}}
%\label{default}
%\end{center}
%\end{figure}

Fig. \ref{Fig1_core}(c) shows the wide-energy photoemission spectra of TlNi$_{2}$Se$_2$. Clear double-peak features due to the spin-orbit interaction are identified for Se $3d$ states at binding energies ($E_B$) = 54.0 and 54.9 eV, Tl $5d$ states at $E_B$ = 12.9 and 15.1 eV, and Ni 3$p$ states with $E_B$ = 66.3 and $E_B$ 67.9 eV (as shown in the inset), confirming the sample composition. In Fig. \ref{Fig1_core}(d), we plot the LDA band structure calculated for the experimental lattice parameters \cite{MHFang_TlNi2Se2_2013} and z$_{Se}$=0.355. Four different bands are crossing the Fermi level, and the band structure exhibits non-negligible three-dimensionality. The density of states (DOS) near $E_F$ is contributed mainly by the Ni $3d$ states with a partial Se $4p$ spectral weight, as seen from Fig. \ref{Fig1_core}(d) (the Se $4p$ spectral weight is indicated by the line-width), and the calculated DOS in Fig. \ref{Fig1_core}(e). 
We notice that one specific band, $\gamma$, shows a back-bending camelback-shaped feature at the Z point (as shown in the dashed-line circle in Fig. \ref{Fig1_core}(d)), leading to a flat band top sitting very close to $E_F$ and contributing to a sharp peak in the DOS (inset of Fig. \ref{Fig1_core}(e)). This gives us a first indication that the heavy electrons reported in TlNi$_{2}$Se$_2$ may originate from this flat band top.
This band results from the hybridization of the Ni-$d_{xy}$ and Se-$p{_z}$ orbitals and is a universal feature in transition metal pnictides and chalcogenides. The particular shape of its dispersion, with a back-bending feature near the $\Gamma$ and Z points, is due to the combination of two opposite binding effects. Indeed, while the interaction of Ni-$d_{xy}$ with Se-p${_z}$ is antibonding and causes the band to disperse upwards toward Z because of an increasing hybridization, the nearest-neighbor Ni-$d_{xy}$ -- Ni-$d_{xy}$ interaction becomes bonding near this point, which explains the energy minimum. In most iron pnictides, the resulting saddle points are found in the unoccupied states \cite{Andersen_Boeri_Calculation}, but the originality of TlNi$_{2}$Se$_{2}$ is that the Fermi level crosses it.

\begin{figure}[!t]
\begin{center}
\includegraphics[width=3.4in]{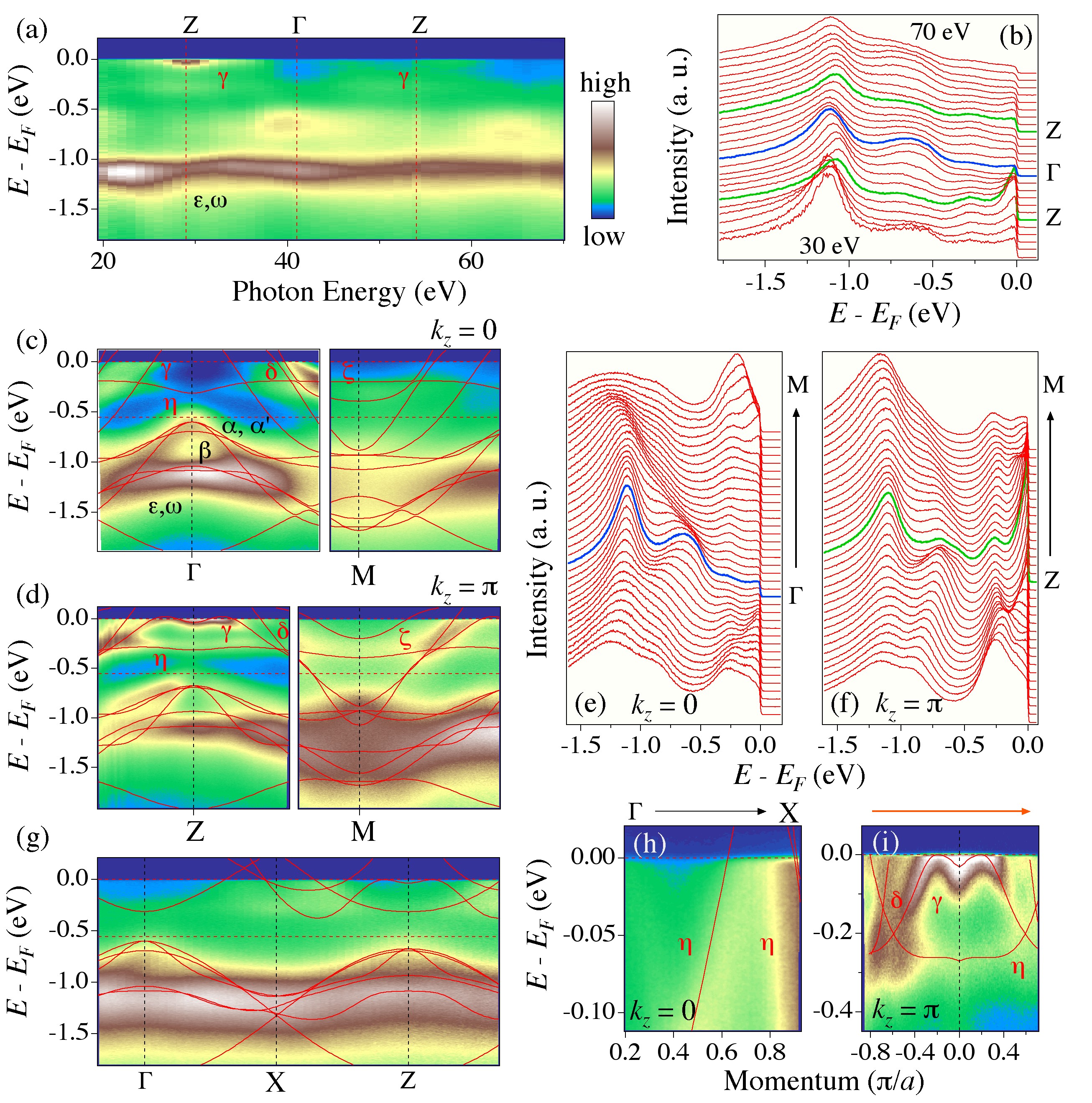}
\end{center}
\caption{\label{Fig2_BS}(Color online) (a) Photon energy dependent ARPES intensity plot recorded at normal emission. (b) Corresponding EDC plot. (c)-(d) ARPES intensity plot along the $\Gamma$-M and Z-A directions. LDA results renormalized by a factor 1.4 are overlapped. (e)-(f) Corresponding EDC plots. (g) ARPES intensity plot along the $\Gamma$-X-Z momentum path, with renormalized LDA. (h) Near-E$_F$ zoom along the $\Gamma$-X direction. (i) Same as (h), with the location in k-space illustrated by the yellow arrow in Fig. \ref{Fig3_FS}(c).
}
\end{figure}

To fully explore the valence band structure in the three dimensional $k$-space, we performed ARPES experiments over a wide photon-energy range to cover more than one BZ in the $k_z$ direction. In Figs. \ref{Fig2_BS}(a)-\ref{Fig2_BS}(b), we plot normal emission ARPES intensities as a function of photon energy and their corresponding energy distribution curves (EDCs). Clear periodic variations of peak positions, especially for the $\gamma$ band near $E_F$ and the $\epsilon$/$\omega$ band at $E_B$ $\sim$ 1.2 eV, are observed with tuning the $k_z$ value by changing the photon energy, indicating strong three dimensionality of the band structure in TlNi$_2$Se$_2$. Using the nearly-free electron approximation with an inner potential of 17 eV, we estimate $h\nu$ = 34 eV for the $k_z$ = 0 plane, and $h\nu$ = 29/54 eV for the $k_z$ = $\pi$ planes. 
We plot the ARPES intensity along the $\Gamma$-M and Z-M directions in Figs. \ref{Fig2_BS}(c)-\ref{Fig2_BS}(d), which are recorded with $h\nu$ = 34 eV and 29 eV, respectively. The corresponding EDCs are shown in Figs. \ref{Fig2_BS}(e)-\ref{Fig2_BS}(f). At first sight, the overall band structure of TlNi$_2$Se$_2$ shows great similarity with the iron pnictide BaFe$_2$As$_2$ \cite{Ding_CMJP} and cobalt pnictide BaCo$_2$As$_2$ \cite{Xu_Co}. The $\alpha, \alpha'$ and $\beta$ hole-like pockets observed at the BZ center ($\Gamma$ point) in BaFe$_2$As$_2$ are fully filled in TlNi$_2$Se$_2$, with the bands topping $\sim$ 0.6 eV below $E_F$. The $\gamma$ and $\delta$ electron-like pockets located at the M point in BaFe$_2$As$_2$ are more than half filled in TlNi$_2$Se$_2$ and form hole-like pockets at the $\Gamma$ point. One flat band ($\eta$) sits $\sim$ 250 meV below $E_F$ in TlNi$_2$Se$_2$, which is also observed near $E_F$ in BaCo$_2$As$_2$ \cite{Xu_Co,Raj_BaCo2As2}. Besides all these $3d$ states, the additional $\zeta$ band is observed in TlNi$_2$Se$_2$ around the M point. Our ARPES data suggest that TlNi$_2$Se$_2$ shares a universal band structure with BaFe$_2$As$_2$ and BaCo$_2$As$_2$, with a chemical potential shifted due to more $3d$ electrons in TlNi$_2$Se$_2$ ($3d^{8.5}$) than in BaFe$_2$As$_2$ ($3d^{6}$) and BaCo$_2$As$_2$ ($3d^{7}$). 
This is indeed what is expected from our LDA calculations, which match the experimental data very well with a renormalization factor of 1.4 as seen in Figs. \ref{Fig2_BS}(c)-\ref{Fig2_BS}(d). 
TlNi$_2$Se$_2$ can thus be viewed as heavily electron doped TlFe$_2$Se$_2$ ($3d^{6.5}$), as we will discuss more later.

\begin{figure}[!t]
\begin{center}
\includegraphics[width=3in]{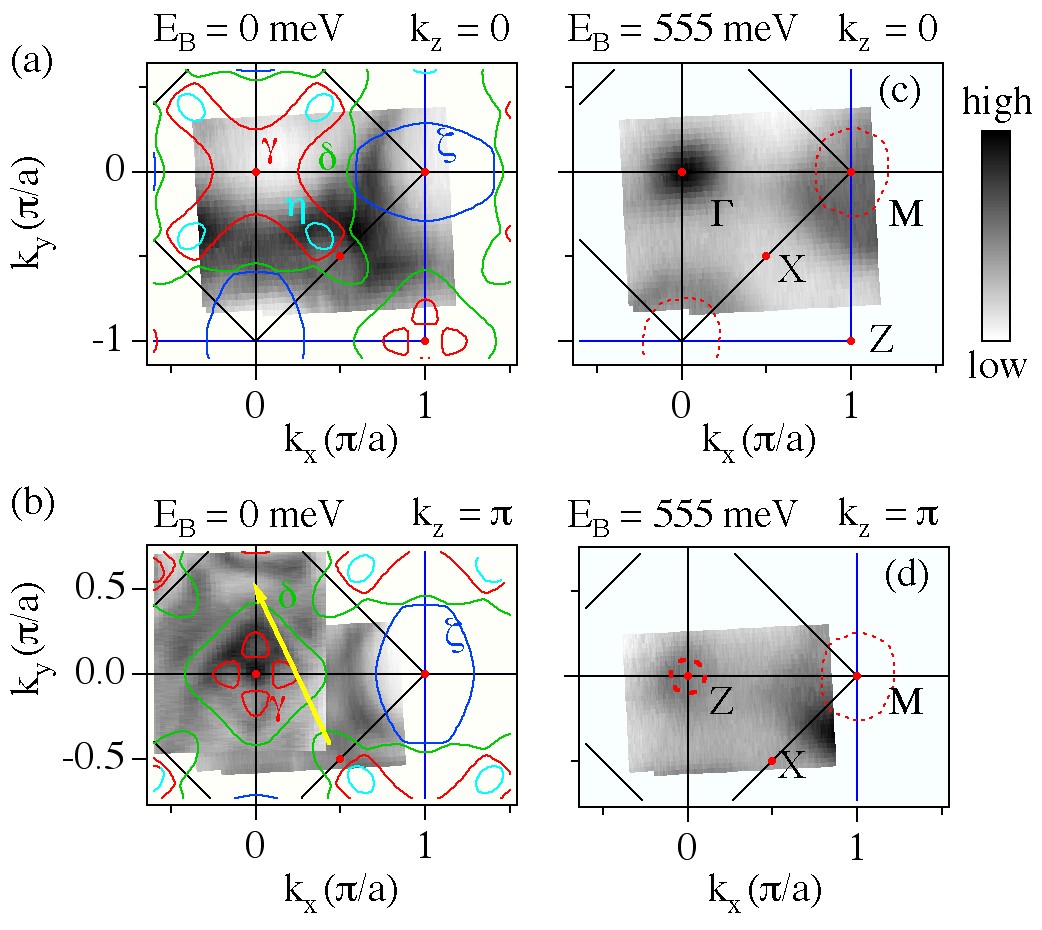}
\end{center}
\caption{\label{Fig3_FS}(Color online) (a)-(b) FS mappings for the $k_z$ =0 and $\pi$ planes, respectively, obtained by integrating the ARPES intensity within $E_F$ $\pm$ 5 meV. (c)-(d) same as (a)-(b) but recorded 555 meV below $E_F$.
}
\end{figure}

In Fig. \ref{Fig2_BS}(g), we plot the ARPES intensity along $\Gamma$-X-Z obtained with photon energy $h\nu$ = 54 eV. The overall band structure is also in a good agreement with the renormalized LDA calculations. In order to examine the details for the bands near $E_F$, high resolution measurements have been recorded along $\Gamma$-X at the $k_z$ = 0 plane ($h\nu$ = 41 eV). As shown in Fig. \ref{Fig2_BS}(h), the hole-like $\eta$ band crosses $E_F$, with the intensity enhanced due to the matrix element effect. Similarly, Fig. \ref{Fig2_BS}(i) shows the band structure in the $k_z$ = $\pi$ plane ($h\nu$ = 29 eV), cutting slightly off the Z-X direction (as illustrated by the yellow arrow in Fig. \ref{Fig3_FS}(c)), in order to well separate the $\gamma$ and $\delta$ bands. In agreement with LDA, the $\gamma$, $\delta$ and $\eta$ bands are observed, hybridizing with each other at their intersections.

\begin{figure*}[!t]
\begin{center}
\includegraphics[width=7in]{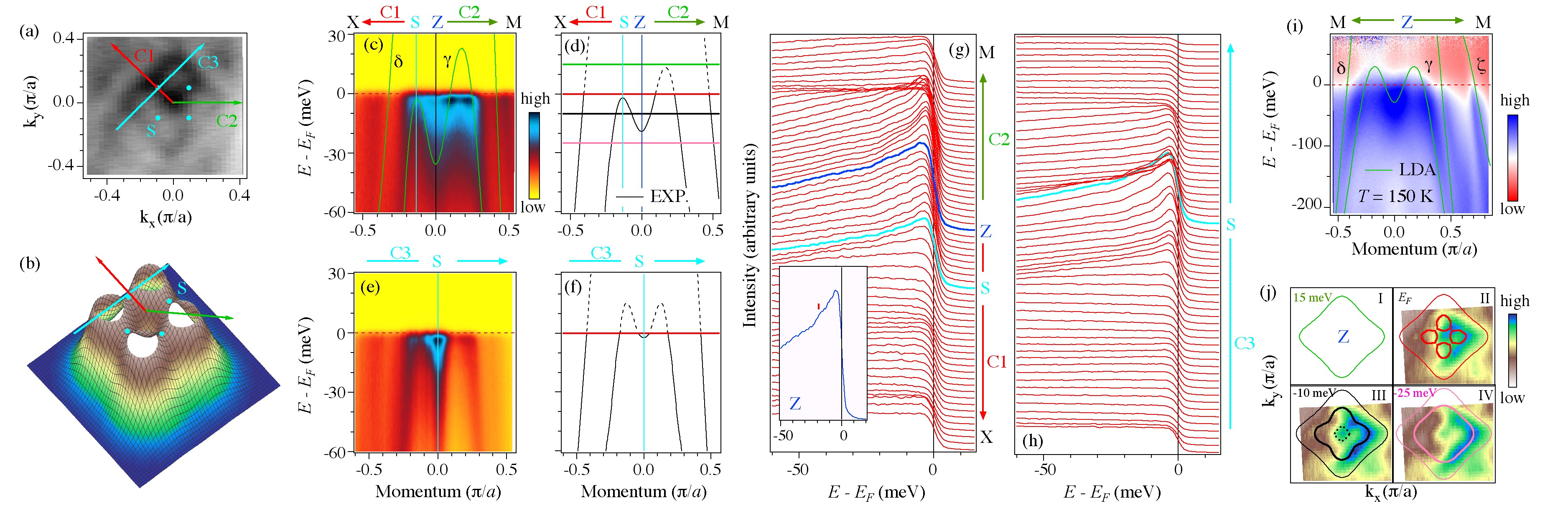}
\end{center}
\caption{\label{Fig4_ZP}(Color online) (a) FSs plot near the Z point. (b) Three-dimensional band structure plot near the Z point. (c) High resolution ARPES results along C1-C2 in (a), recorded at $T$ = 5 K. Renormalized LDA shifted by 7 meV are overlapped, also in (i). (d) Corresponding extracted band dispersion. (e) ARPES results along C3 in (a). (f) Corresponding extracted band dispersion. (g)EDC plots for C1-C2. The inset is the EDC taken at the Z point. (h) EDC plots for C3. (i) High temperature ARPES measurements along M-Z-M at $T$ = 150 K. (e) Constant energy map contours at 15 meV above $E_F$, $E_F$, 10 and 25 meV below $E_F$. The corresponding constant energy ARPES intensity plots are also overlapped.}
\end{figure*}

In Figs. \ref{Fig3_FS}(a) and \ref{Fig3_FS}(b), we display the Fermi surface at $k_z$ = 0 and $\pi$, respectively. The corresponding LDA calculations (slightly shifted upward by 7 meV) are also overlapped. As expected from the previous analysis of the band structure in the vicinity of $E_F$, the FS topology of TlNi$_2$Se$_2$ is quite different from that of Tl(K)$_x$Fe$_{2-y}$Se$_2$. The $\gamma$ and $\delta$ bands, which form a pair of degenerate electron-like pockets at the M point in Tl(K)$_x$Fe$_{2-y}$Se$_2$\cite{Qian_PRL2011,Y_Zhang_NatureMat2011, D_MouPRL2011, XP_WangEPL2011}, become hole-like pockets crossing $E_F$ at the $\Gamma$ point. The $\delta$ band crosses the Z point with a smaller $k_F$ and the $\gamma$ band creates four small lobes because of the back-bending feature discussed before. The $\eta$ band forms a small hole-like pocket near the X point, and the $\zeta$ band forms an electron-like pocket at the M point. In contrast, we find that the constant energy maps recorded at 550 meV below $E_F$ for the $k_z$ = 0 and $\pi$ planes, which are, respectively, displayed in Figs. \ref{Fig3_FS}(c) and  \ref{Fig3_FS}(d), are quite similar to the FSs of Tl(K)$_x$Fe$_{2-y}$Se$_2$, on which superconducting gaps are observed below $T_c$ \cite{XP_WangEPL2011}. To illustrate this resemblance, we overlap on these figures the FSs obtained for Tl(K)$_x$Fe$_{2-y}$Se$_2$ in Ref. \cite{XP_WangEPL2011}. Similarly to the case of BaFe$_2$As$_2$ and BaCo$_2$As$_2$, TlNi$_2$Se$_2$ can be interpreted as a heavily electron doped TlFe$_2$Se$_2$ as a first approximation. Therefore, our results suggest that TlFe$_2$Se$_2$, the phase without vacancy ordering, is the superconducting compound of the iron-chalcogenide Tl$_x$Fe$_{2-y}$Se$_2$, which is consistent with the STM results \cite{W_LiNatPhys8} and supports the conclusions derived from previous ARPES experiments on AFe$_{2-x}$Se$_2$  \cite{Qian_PRL2011,Y_Zhang_NatureMat2011, D_MouPRL2011, XP_WangEPL2011, ZH_LiuPRL2012, XP_WangEPL2012}.

The significant reduction of electron correlation strength in TlNi$_2$Se$_2$ (as indicated by a small renormalization factor of 1.4) as compared to TlFe$_2$Se$_2$ (renormalization factor of 2.5 \cite{Qian_PRL2011}) is anticipated due to the increased filling of the electronic $3d$ shell. Indeed, in Ref. \cite{Xu_Co} it was argued that the stronger correlations in BaFe$_2$As$_2$ as compared to BaCo$_2$As$_2$ are driven by the lower band filling in the presence of strong Hund's coupling.
The mass enhancement due to correlation effects in TlNi$_2$Se$_2$ is much smaller than that deduced from the specific heat coefficient by assuming 1.5 carriers/Ni and a spherical Fermi surface \cite{MHFang_TlNi2Se2_2013}. Here we suggest the large specific heat coefficient to be related to the flat band near the chemical potential, as unveiled by our DFT calculations and the ARPES data (Fig. \ref{Fig2_BS}(d)).
To check the fine details of this $\gamma$ band top, we performed high resolution measurements at $T$ = 5 K, above $T_c$. As seen from the result in Fig. \ref{Fig4_ZP}(c) and the corresponding EDC plot in Fig. \ref{Fig4_ZP}(g), the top of the $\gamma$ band sits at the S point, $\sim$ 2 meV below the $E_F$ along the Z-X direction (C1 in Fig. \ref{Fig4_ZP}(a)), and bends back with the band bottom at $\sim$ 18 meV below the $E_F$ (indicated by the EDC taken at the Z point shown in the inset of Fig. \ref{Fig4_ZP}(g)). Along the Z-M direction (C2 in Fig. \ref{Fig4_ZP}(a)), the band top of $\gamma$ is slightly above $E_F$, leading the $\gamma$ band to double-cross $E_F$ and forming small lobes. The $\gamma$ band top along Z-M is estimated at $\sim$ 15 meV above $E_F$ from the high temperature data (150 K) divided by the Fermi function shown in Fig. \ref{Fig4_ZP}(i). 
In Fig. \ref{Fig4_ZP}(e), we plot the ARPES intensity along the momentum path passing through the S point and perpendicular to the Z-X direction (C3 in Fig. \ref{Fig4_ZP}(a)). The corresponding EDC plot is also shown in Fig. \ref{Fig4_ZP}(h). The $\gamma$ band crosses $E_F$ four times (passing through two small hole-like lobes), and has a negative curvature at the S point along C1, as seen from Fig. \ref{Fig4_ZP}(e), and the extracted band structure in Fig. \ref{Fig4_ZP}(f). At the same time, the $\gamma$ band has a positive curvature at the S point along the perpendicular direction C1 as shown in Figs. \ref{Fig4_ZP}(c)-\ref{Fig4_ZP}(d). Therefore, our data confirms the picture discussed before and in particular the camelback shape and the VHS at the S points. This can be better visualised in the band structure plotted in Fig. \ref{Fig4_ZP}(b). In fact, this VHS is guaranteed by the asymmetry of the back-bending feature as shown in Fig. \ref{Fig4_ZP}(c) and the extracted band structure in Fig. \ref{Fig4_ZP}(d) along the X-Z-M direction.
The existence of the four flat parts of the dispersion very close to $E_F$ provides a natural explanation for the specific heat coefficient in this weakly correlated system TlNi$_2$Se$_2$.

Because the VHS is very close to $E_F$, the FS topology of the $\gamma$ band changes dramatically when slightly tuning the chemical potential. When we put the chemical potential $>$ 15 meV above $E_F$, only the $\delta$ band crosses at the Z point and forms a hole-like pocket (Fig. \ref{Fig4_ZP}(e) I). By putting the chemical potential at -15 meV $<$ E$_B$ $<$ 2 meV (for instance the red lines in Fig. \ref{Fig4_ZP}(d)), the $\gamma$ band crosses the chemical potential but only along the Z-M directions, forming four small hole-like lobes (Fig. \ref{Fig4_ZP}(e) II). When the chemical potential is setting in the range of 2 meV$<E_B<$17 meV, the $\gamma$ band crosses the chemical potential twice both along the Z-M and Z-X directions, forming a small electron-like pocket inside a concentric hole-like pocket (Fig. \ref{Fig4_ZP}(e) III). Further shifting the chemical potential down to $E_B >$17 meV, the inner electron pocket disappears, leaving two hole-like pockets at the Z point (Fig. \ref{Fig4_ZP}(e) IV). All these Lifshitz transitions associated with VHS happen with the chemical potential varying by less than 30 meV, providing a possible explanation for why $T_c$ of KNi$_2$Se$_2$ ($T_c \sim$ 0.8 K) \cite{2012_KNi2Se2_SC} is $\sim$ 4 times smaller than that of TlNi$_2$Se$_2$. 
Because K is much more sensitive to air than Tl, a slight shift of chemical potential in KNi$_2$Se$_2$ is expected. The difference of DOS near $E_F$ due to this small chemical potential variation cannot explain $\sim$ 4 times smaller $T_c$ in KNi$_2$Se$_2$. On the other hand, the FS topology change associated with VHS due to the chemical potential shift is one candidate interpretation for the large variation of superconductivity. Most interestingly, since the strength of the hybridization of Ni-$d_{xy}$ with the Se-$p_{z}$ orbital controls the position of the saddle point with respect to the Fermi level, the shape of the Fermi surface is highly sensitive to the height of the selenium atom over the Ni-Se plane. This may be related to the sudden $T_c$ drop under pressure recently reported in TlNi$_2$Se$_2$ \cite{2014_NiSe122_pressure}.

It is also worth mentioning that the end-member of the hole-doped BaFe$_2$As$_2$, the iron pnictide KFe$_2$As$_2$, shows some similarities with TlNi$_2$Se$_2$. First, a large specific heat coefficient $\sim$ 94 mJ/mol K$^2$ is also reported in KFe$_2$As$_2$ \cite{KFe2As2_specificheat}. Also, a similar pressure effect has been observed in KFe$_2$As$_2$ \cite{Tafti_NPhys2013} and TlNi$_2$SeS\cite{2014_NiSe122_pressure}. Finally, a flat band bottom for the $\beta$ band forming four small hole-like lobes at the M point due to hybridization with an electron band, has also been observed by ARPES \cite{Sato_KFe2As2,Xu_OD}. And recently a VHS \cite{2014_KFA_VHS} has been observed and suggested to be responsible for the nodal superconductivity reported in this material \cite{ReidPRL109,Tafti_NPhys2013,TanatarPRL2010}.

This work was supported by the Swiss National Science Foundation (Grant No. 200021-137783), the Sino-Swiss Science and Technology Cooperation (Project No. IZLCZ2138954), and MOST (Grant No. 2010CB923000, 2011CBA001000, 2011CBA00102, 2012CB821403 and 2013CB921703) and NSFC (11004232, 11034011/A0402, 11234014 and 11274362) from China, tthe National Science Foundation under Grant No. NSF PHY11-25915, IDRIS/GENCI Orsay under project 091393, and the European Research Council under project 617196. The work in ZJU was supported by the Natural Science Foundation of China (Grants No. 11374261 and No. 11204059), and the National Basic Research  Programs of China (No. 2011CBA00103, No. 2012CB821404, and No. 2015CB921004).

%\bibliography{biblio_en}
\bibliography{biblio_ens}

\begin{thebibliography}{31}
\bibitem [{ram()}]{raman_exp}%
  \BibitemOpen
  \href@noop {} {\bibinfo  {journal} {{Here x$'$ and y$'$ are parallel to the
  Fe-Fe bounding while x and y are at 45$^{\circ}$ from these axes. The inset
  shows the evolution of the Raman intensity of the $B_{1g}$ mode as a function
  of the angle between the incident polarization $e_i$ and the x axis, for
  $e_i$ parallel and perpendicular to the scattered light polarization $e_s$}}\
  }\BibitemShut {NoStop}%
\end{thebibliography}

\end{document}